\documentclass{article} % For LaTeX2e
\usepackage{iclr2025_conference,times}

\usepackage{amsmath,amsfonts,bm}

\def\eqref#1{equation~\ref{#1}}
\def\1{\bm{1}}

\DeclareMathAlphabet{\mathsfit}{\encodingdefault}{\sfdefault}{m}{sl}
\SetMathAlphabet{\mathsfit}{bold}{\encodingdefault}{\sfdefault}{bx}{n}

\usepackage{hyperref}
\usepackage{url}
\usepackage{color}
\usepackage{graphicx}
\usepackage{booktabs}
\usepackage{multirow}
\usepackage{authblk}

\title{\textsc{FinDER}: Financial Dataset for Question Answering and Evaluating Retrieval-Augmented Generation}

\author[1,*]{Chanyeol Choi}
\author[1,*]{Jihoon Kwon}
\author[1]{Jaeseon Ha}
\author[1]{Hojun Choi}
\author[1]{Chaewoon Kim}
\author[2]{Yongjae Lee}
\author[3]{Jy-yong Sohn}
\author[4,*]{Alejandro Lopez-Lira}

\affil[1]{LinqAlpha}
\affil[2]{UNIST}
\affil[3]{Yonsei University}
\affil[4]{University of Florida}
\affil[*]{Corresponding Authors}
\iclrfinalcopy % Uncomment for camera-ready version, but NOT for submission.
\begin{document}

\maketitle

\begin{abstract}
In the fast-paced financial domain, accurate and up-to-date information is critical to addressing ever-evolving market conditions. Retrieving this information correctly is essential in financial Question-Answering (QA), since many language models struggle with factual accuracy in this domain. We present \textsc{FinDER}, an expert-generated dataset tailored for Retrieval-Augmented Generation (RAG) in finance. Unlike existing QA datasets that provide predefined contexts and rely on relatively clear and straightforward queries, \textsc{FinDER} focuses on annotating search-relevant evidence by domain experts, offering 5,703 query–evidence–answer triplets derived from real-world financial inquiries. These queries frequently include abbreviations, acronyms, and concise expressions, capturing the brevity and ambiguity common in the realistic search behavior of professionals. By challenging models to retrieve relevant information from large corpora rather than relying on readily determined contexts, \textsc{FinDER} offers a more realistic benchmark for evaluating RAG systems. We further present a comprehensive evaluation of multiple state-of-the-art retrieval models and Large Language Models, showcasing challenges derived from a realistic benchmark to drive future research on truthful and precise RAG in the financial domain.
\end{abstract}

\section{Introduction}

Accurate information retrieval is critical in financial Question-Answering (QA)~\citep{setty2024improving, iaroshev2024evaluating, sarmah2023towards}, where even small errors can lead to costly consequences in investments, risk management, and compliance~\citep{gozman2014role, hopkin2018fundamentals}. However, ensuring precision is increasingly difficult due to the dynamic and complex nature of financial data~\citep{liu2024dynamic, frischbier2020managing}. With new information constantly being updated, retrieval systems face challenges in navigating vast documents, dense tables, and context-dependent narratives from sources like financial reports and market feeds~\citep{so2022assessing, jiang2014structure}. Moreover, financial queries are often brief, ambiguous, and filled with domain-specific jargon and abbreviations~\citep{banks2004financial, downes2014dictionary, law2014dictionary} (e.g., “Recent CAGR in MS trading revenue”), requiring systems to first identify key contextual elements—such as the company name, its business focus, and the specific metrics mentioned—while retrieving the correct evidence. Unlike open-domain QA, financial QA demands a higher level of precision, disambiguation, and technical understanding, which makes it uniquely challenging and error‑prone~\citep{chen2021finqa, zhu2021tat, chen2022convfinqa, zhao2022multihiertt, saini2023evolution}.

Even state-of-the-art Large Language Models (LLMs) struggle with factual correctness in financial queries without proper context~\citep{islam2023financebench, reddy2024docfinqa, chen2024fintextqa, xu2024fintruthqa}. For example, GPT-4-turbo~\citep{achiam2023gpt} achieved only 9\% accuracy when answering clear and straightforward questions in a closed-book setting, with 91\% of its responses being incorrect or unanswered~\citep{islam2023financebench}. These results highlight the importance of providing relevant information to LLMs for accurate performance. However, simply extending context windows by feeding entire financial documents into LLMs has proven ineffective due to computational cost and processing latency~\citep{li2024long, li2024retrieval, wang2024beyond}. Thus, relying solely on LLMs is insufficient for finance-specific tasks. This is where Retrieval-Augmented Generation (RAG)~\citep{lewis2020retrieval} becomes essential. By searching and pinpointing relevant information within large financial documents efficiently and feeding it to LLMs, RAG pipelines ensure accurate, explainable answers that meet the precision demands of financial QA~\citep{setty2024improving, iaroshev2024evaluating}.

However, prior datasets~\citep{de2018inf, chen2021finqa, zhu2021tat, chen2022convfinqa, zhao2022multihiertt, islam2023financebench, reddy2024docfinqa, chen2024fintextqa, xu2024fintruthqa} that rely on structured questions with readily available context have failed to reflect the importance of ambiguous queries and the retrieval process, which is central to financial QA. To address these limitations, we introduce \textsc{FinDER} (\textbf{Fin}ancial \textbf{D}ataset for \textbf{E}valuating \textbf{R}AG), a dataset specifically designed to capture the ambiguity and context-dependency of real-world financial queries. \textsc{FinDER} captures this complexity by sampling real search queries from professionals in financial service, with financial experts linking each query to ground-truth evidence extracted from a company’s annual report (10-K) filings and providing carefully verified answers. By focusing on ambiguous, realistic queries that demand contextual understanding, \textsc{FinDER} offers a rigorous testbed for evaluating retrieval systems and LLMs, pushing them to overcome the limitations of prior datasets and better meet the demands of financial QA.

In summary, our contributions include: (1) the creation of the \textsc{FinDER} dataset with 5,703 expert-annotated QA pairs grounded in 10-K reports, focusing on ambiguous query understanding and accurate retrieval; (2) an analysis of \textsc{FinDER}’s characteristics compared to prior datasets, demonstrating its uniqueness in query brevity, use of acronyms, and broad coverage of financial topics; and (3) baseline evaluations of both state-of-the-art retrieval models and LLMs on \textsc{FinDER}. By revealing the strengths and limitations of current approaches and providing a benchmark for future improvements, we offer a challenging testbed for developing more robust retrieval-augmented financial QA systems.

\section{Related Work}

\subsection{Financial Question-Answering Datasets}
Recent years have witnessed the rapid evolution of benchmark datasets for financial question answering (QA), each addressing unique challenges within the domain. Early datasets, such as FiQA~\citep{de2018inf} introduced tasks involving opinion-based QA, while FinQA~\citep{chen2021finqa}, and TAT-QA~\citep{zhu2021tat}, focused on numerical reasoning, and hybrid reasoning across textual and tabular data. However, most existing datasets aim to reflect realistic settings but either neglect retrieval or implement it under limited conditions~\citep{sarmah2024hybridrag}. ConvFinQA~\citep{chen2022convfinqa} and MultiHiertt~\citep{zhao2022multihiertt} focus on conversational queries and multi-table reasoning but do not treat retrieval as a core task, limiting their applicability in real-world search scenarios. DocFinQA~\citep{reddy2024docfinqa} limits retrieval to a single pre-selected relevant document, while FinanceBench~\citep{islam2023financebench} offers limited scalability with only 150 public questions and minimal emphasis on retrieval. FinTextQA~\citep{chen2024fintextqa} aims to address retrieval in open-book settings, but its impact is restricted because the dataset is currently unavailable for public use.

\begin{figure}[ht]
    \centering
    \includegraphics[width=\linewidth]{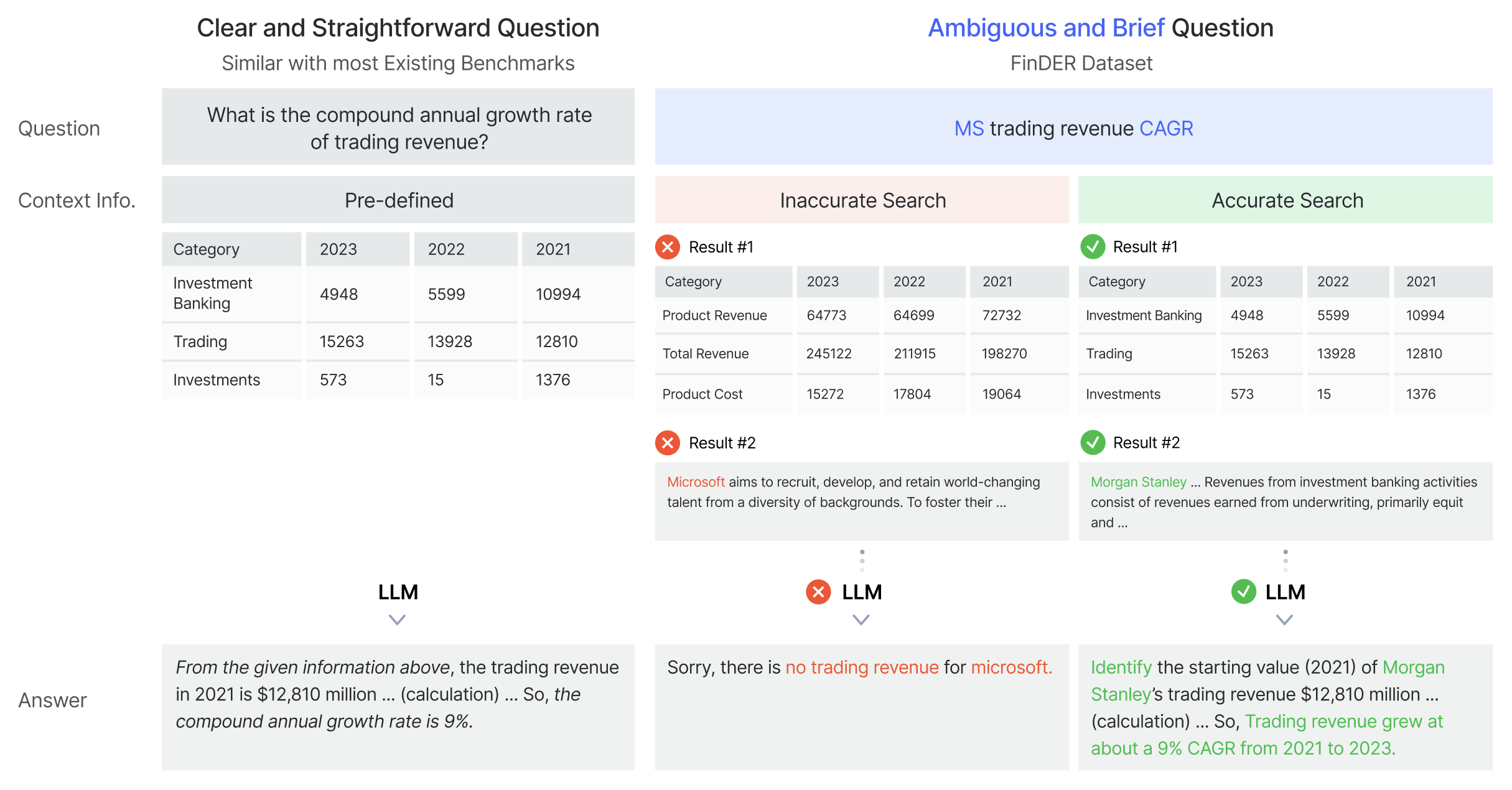}
    \caption{This figure contrasts traditional datasets with predefined context and clear questions against \textsc{FinDER}, which evaluates models on \emph{ambiguous and brief queries that require retrieval}. Unlike existing benchmarks, \textsc{FinDER} uniquely assesses both the search system’s ability to interpret queries (e.g., recognizing ‘MS’ as Morgan Stanley) and the LLM’s capacity to synthesize relevant information from multiple sources to generate accurate responses (e.g., extracting trading revenue data to compute CAGR).}
    \label{fig:teasor}
\end{figure}

\subsection{Retrieval-Augmented Generation (RAG) in Finance}
RAG~\citep{lewis2020retrieval} is a technique designed to improve performance of LLMs by retrieving and integrating relevant external documents into the response generation process to provide contextually rich and reliable outputs~\citep{jiang2023active, gao2023retrieval}. RAG has effectively addressed key limitations of LLMs in finance~\citep{zhang2023enhancing, sarmah2023towards, zhao2024optimizing, setty2024improving, iaroshev2024evaluating, darji2024enhancing}. To be specific, while LLMs excel at natural language tasks, they often produce hallucinated~\citep{huang2023survey, ji2023survey, rawte2023survey, saparov2023testing} or outdated responses due to a lack of up-to-date, domain-specific knowledge~\citep{sun2023head, kandpal2023large, szymanski2024limitations, jayakumar2023large, mai2024llms}—a challenge that is particularly critical in finance where information evolves rapidly. By retrieving relevant documents (e.g., news articles, filings, and knowledge bases), RAG can mitigate the limitations of LLMs, improving both accuracy and contextual richness~\citep{setty2024improving, iaroshev2024evaluating, zhang2023enhancing}. Improving the pipeline of data collection, document indexing, retrieval, and generation~\citep{gao2023retrieval} is crucial for enhancing accuracy and minimizing hallucinations in financial QA systems. By integrating diverse data sources~\citep{zhang2023enhancing}, using effective document chunking~\citep{yepes2024financialreportchunkingeffective}, and leveraging embedding-based retrieval with reranking~\citep{zhao2024optimizing, sarmah2023towards}, RAG ensures precise and contextually relevant inputs for LLMs in financial tasks. By leveraging RAG, LLMs can provide financial professionals with timely, evidence-based insights, thereby enhancing decision-making processes and fostering greater trust.

\section{\textsc{FinDER} Dataset}
\noindent\textbf{Notice for Readers.} Dataset is available in Huggingface\footnote{\url{https://huggingface.co/datasets/Linq-AI-Research/FinDER}}. 

% What is the dataset and why is it needed?
\textsc{FinDER} is a benchmark dataset designed to support financial question answering, comprising 5,703 expert-annotated query–evidence–answer triplets. Unlike existing QA datasets that rely on predefined contexts, \textsc{FinDER} captures the ambiguity and brevity inherent in real-world financial search queries, making it a more representative resource for financial information retrieval and reasoning (See Table~\ref{tab:retrieval_compare_wellformed} for detailed comparison).

\textsc{FinDER} consists of four key components:

\begin{itemize}
    \item \textit{Documents} – A collection of annual reports, serving as the primary source of financial information.
    \item \textit{Questions} – A set of expert-annotated financial inquiries reflecting real-world search behavior in the financial domain.
    \item \textit{Ground truth evidences} – One or more passages from the document set that are manually selected to contain the necessary information for answering each question.
    \item \textit{Answers} – Labeled responses that represent the correct information retrievable from the corresponding evidence.
\end{itemize}

By structuring \textsc{FinDER} with these four components, the dataset enables comprehensive evaluation of both retrieval and generation tasks, making it a valuable resource for advancing RAG development.

\subsection{Data Collection}
\label{3_data_collection}
% Where did the data come from and how was it collected?
\textsc{FinDER} is constructed using real-world financial inquiries from investment professionals, ensuring its relevance to industry applications. The dataset covers companies from the S\&P 500 index as of December 31, 2024, and is built upon two primary data sources: a document set of annual reports and a set of expert-annotated questions. The \textit{documents} consist of the latest Form 10-K filings, which were collected via web scraping from EDGAR\footnote{\url{https://www.sec.gov/edgar/search/}} in raw HTML format. The \textit{questions} were initially gathered from a financial Q\&A service database used by hedge fund analysts, portfolio managers, and investment banking analysts. To ensure diversity and relevance, duplicate queries were removed and a balanced sampling across S\&P 500 companies was applied. From an initial collection of 7,000 questions, we applied a rigorous filtering process: any question for which no ground truth evidence could be identified in the corresponding 10-K filing was excluded. Similarly, companies for which no questions were associated were removed from the dataset. This filtering resulted in a final dataset comprising 5,703 questions linked to reports from 490 companies. This refined structure makes \textsc{FinDER} a robust and representative benchmark for evaluating financial question answering systems.

\subsection{Annotation Process}
\label{3_annotation}
To ensure high-quality mappings between queries, supporting evidence, and answers in financial question-answering, we adopt a meticulously designed, multi-stage annotation process that leverages the expertise of financial domain professionals. This structured approach guarantees accuracy, relevance, and consistency in extracting insights from financial reports. The annotation process is conducted by two domain experts: an investment bank analyst and a Certified Public Accountant (CPA). Before initiating the annotation process, they receive detailed guidelines emphasizing the following principles:

\textbf{Ground Truth Evidence Relevance:} Annotators are required to identify the most pertinent sections—such as paragraphs, tables, or figures—within 10-K filings that directly address the given query.

\textbf{Answer Generation and Verification:} Responses have to be formulated with accurate and precisely with calculations if needed, ensuring that they were both comprehensive and strictly grounded in the extracted evidence.

The annotation process follows a rigorous, cross-validated framework designed to minimize errors and enhance reliability. The methodology consists of several distinct stages: First, annotators independently review the relevant 10-K filings to select candidate evidence snippets that directly addressed the query\footnote{We perform basic preprocessing by converting HTML files into plain text, removing HTML tags, and segmenting the content into distinct paragraphs. This process ensures a structured and well-organized format for annotation.}. This step ensures a multi-perspective selection of supporting evidence, reducing bias and improving coverage. Next, based on the identified evidence, they formulated initial answers that are clear, precise, and entirely derived from authoritative financial documents. To maintain consistency across responses, these draft answers undergo standardization using LLM (GPT-o1~\citep{jaech2024openai}), ensuring a uniform format while preserving expert judgment. Finally, the process incorporated a cross-validation and refinement phase, where annotators mutually reviewed each other's work. Any discrepancies in evidence selection or answer formulation were discussed and resolved collaboratively, ensuring that the final dataset accurately reflected the content of the 10-K filings.

By integrating multiple expert perspectives, structured cross-validation, and systematic quality control, the annotation process ensures that financial question-answering annotations are both highly accurate and well-grounded in authoritative sources. This meticulous approach guarantees consistency, minimizes inaccuracies, and enhances the overall reliability of the dataset for financial analysis and decision-making.

\subsection{Statistics}
\label{3_statistics}
The \textsc{FinDER} dataset is designed to reflect the way financial professionals search for information, incorporating both domain-specific expressions and a diverse set of financial questions. The dataset ensures that models trained on it must handle real-world complexities, including specialized terminology, numerical reasoning, and various aspects of financial disclosures.

\begin{figure}[h]
    \centering
    \includegraphics[width=0.6\linewidth]{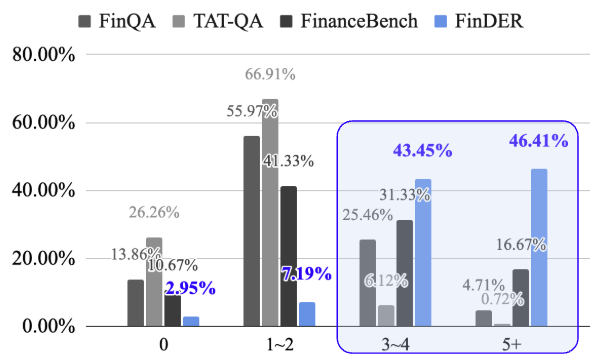}
    \caption{Comparison of the number of domain-specific expressions (jargon, abbreviations, acronyms) used in questions across different benchmarks. FinDER contains a significantly higher proportion of questions with a large number of domain-specific expressions (3+), with 43.45\% in the 3 to 4 range and 46.41\% in the 5+, surpassing other benchmarks.}
    \label{fig:statistics_abbreviations}
\end{figure}

A key feature of \textsc{FinDER} is its extensive use of financial jargon, abbreviations, and acronyms. As shown in Figure~\ref{fig:statistics_abbreviations}, a significant proportion of queries contain multiple domain-specific expressions, with 43.45\% falling within the 3–4 expression range and 46.41\% containing five or more specialized terms~\footnote{We analyzed the number of domain-specific expressions using \texttt{GPT-4o-mini}~\citep{achiam2023gpt}.}. This demonstrates that the dataset effectively captures the natural search behavior of financial analysts, who rely heavily on precise terminology when querying 10-K reports. Unlike general-purpose question-answering datasets~\citep{chen2021finqa, zhu2021tat, islam2023financebench}, \textsc{FinDER} requires models to handle a high density of financial-specific vocabulary, making it a challenging and domain-adaptive benchmark.

\begin{table}[h]
    \centering
    \begin{tabular}{lrr}
        \toprule
        \textbf{Question Category} & \textbf{Count} & \textbf{Percentage} \\
        \midrule
        Accounting         & 491  &  8.61\% \\
        Company Overview   & 1081 & 18.95\% \\
        Financials         & 990  & 17.36\% \\
        Footnotes          & 953  & 16.71\% \\
        Governance         & 718  & 12.59\% \\
        Legal              & 490  &  8.59\% \\
        Risk               & 490  &  8.59\% \\
        Shareholder Return & 490  &  8.59\% \\
        \midrule
        \textbf{Total}     & 5703 & 100.00\% \\
        \bottomrule
    \end{tabular}
    \caption{Categorization of questions based on the topics they address in the 10-K report. This table shows the distribution of questions according to the specific aspects of financial disclosures they are related to.}
    \label{tab:question_categories}
\end{table}

Beyond terminology, \textsc{FinDER} encompasses a broad spectrum of financial questions. Table~\ref{tab:question_categories} illustrates the distribution of queries across major financial topics. The dataset includes company overviews (18.95\%), financial statement analysis (17.36\%), governance (12.59\%), and legal disclosures (8.59\%), ensuring comprehensive coverage of the key components found in corporate filings. By incorporating a wide range of question types, the dataset aligns with the diverse information needs of finance professionals, from investors assessing risk to auditors verifying compliance.

In addition to encompassing a wide range of financial topics, \textsc{FinDER} includes both qualitative and quantitative reasoning tasks. As shown in Table~\ref{tab:overall_stats}, 84.52\% of queries require qualitative reasoning, such as interpreting textual information or assessing financial risks, while 15.48\% involve quantitative reasoning, requiring numerical calculations and financial modeling.
Following the quantitative question categorization used in previous work~\citep{zhu2021tat, chen2021finqa}, Table~\ref{tab:quant_breakdown} indicates that a substantial portion (49.83\%) of quantitative queries involve compositional reasoning, where multiple steps are necessary to derive the correct answer. Additionally, operations such as division (14.50\%), multiplication (13.70\%), and subtraction (13.48\%) further emphasize the dataset’s focus on computational financial analysis.

\begin{table}[htbp]
\centering
\begin{minipage}[t]{0.45\linewidth}
    \centering
    \caption{Overall Distribution of Reasoning Types}
    \label{tab:overall_stats}
    \begin{tabular}{lcc}
    \toprule
    \textbf{Reasoning Type} & \textbf{Count} & \textbf{Percentage} \\
    \midrule
    Quantitative & 883   & 15.48\% \\
    Qualitative & 4820   & 84.52\% \\
    \midrule
    \textbf{Total} & 5703 & 100\% \\
    \bottomrule
    \end{tabular}
\end{minipage}
\hspace{1em}
\begin{minipage}[t]{0.45\linewidth}
    \centering
    \caption{Breakdown of Quantitative Reasoning Subcategories}
    \label{tab:quant_breakdown}
    \begin{tabular}{lcc}
    \toprule
    \textbf{Subcategory} & \textbf{Count} & \textbf{Percentage} \\
    \midrule
    Addition       & 75  & 8.49\% \\
    Subtraction    & 119 & 13.48\% \\
    Multiplication & 121 & 13.70\% \\
    Division       & 128 & 14.50\% \\
    Compositional  & 440 & 49.83\% \\
    \midrule
    \textbf{Total} & 883 & 100\% \\
    \bottomrule
    \end{tabular}
\end{minipage}
\end{table}

By integrating domain-specific terminology, a diverse set of financial topics, and a balanced mix of qualitative and quantitative reasoning, \textsc{FinDER} presents a realistic and rigorous benchmark for financial question-answering. The dataset challenges models to not only retrieve relevant financial information but also interpret and reason through complex queries, making it a crucial resource for advancing AI-driven financial analysis.

\section{Experimental Setup}

To evaluate our baseline, we adopt the RAGAS framework~\citep{es2023ragas}, which provides automated evaluation for RAG systems and integrates seamlessly with LLM-based workflows such as \textsc{LangChain}\footnote{\url{https://www.langchain.com/}} and \textsc{LlamaIndex}\footnote{\url{https://www.llamaindex.ai/}}. It offers a suite of metrics covering aspects such as retrieval relevance and generation faithfulness. For analysis, we evaluate our system on a representative 10\% subset of the dataset.

The RAG pipeline involves pre-processing steps including document parsing and indexing~\citep{gao2023retrieval, finardi2024chronicles, singh2024chunkrag, li2025enhancing}, as well as transformations applied to both documents and queries~\citep{efthimiadis1996query, wang2011cascade, carpineto2012survey, nogueira2019document, wang2023query2doc, chan2024rq}. Due to the variability introduced by these steps, rule-based evaluation is often insufficient. To address this, we adopt an LLM-as-a-judge approach~\citep{gu2024survey, zheng2023judging, huang2024empirical} supported by RAGAS, enabling more flexible and context-aware evaluation of generated outputs.

By combining RAGAS with an LLM-based evaluation strategy, we provide a robust and adaptable assessment of our baseline. This setup effectively captures the diverse outcomes of document and query transformations, aligning with the flexibility needed for advancing RAG research.

\subsection{Baseline Systems for Retrieval Models}
\label{baseline_retrieval}

We evaluate the retrieval component of our Retrieval-Augmented Generation (RAG) system using four state-of-the-art models: one sparse and three dense retrievers. For the sparse baseline, we use BM25 with standard parameters \( k_1 = 1.2 \) and \( b = 0.75 \). The dense models include one decoder-based model—\texttt{e5-mistral-7b-instruct} (E5-Mistral)~\citep{wang2023improving}—and two encoder-based models: \texttt{multilingual-e5-large-instruct} (mE5)~\citep{wang2024multilingual}, and \texttt{gte-large-en-v1.5} (GTE)~\citep{li2023towards}.

Our preprocessing pipeline follows a simple approach: raw HTML documents are parsed to remove tags, then segmented into paragraphs, which form the retrieval corpus. For each query, the system retrieves the top-10 paragraphs based on model-specific similarity scores.
We assess performance using the \texttt{Context Recall} metric from RAGAS~\citep{es2023ragas}, which measures how well the retrieved contexts cover reference information. References are decomposed into individual claims, and recall is computed based on whether each claim is supported by the retrieved passages. Following RAGAS, we use LLM-based scoring to estimate recall (and precision), ranging from 0 to 100.

\subsection{Baseline systems for generation models}

We evaluate the generation component of our RAG system using four state-of-the-art language models: \texttt{GPT-o1} from OpenAI~\citep{jaech2024openai}, \texttt{claude-3.7-sonnet}\footnote{\url{https://www.anthropic.com/news/claude-3-7-sonnet}} from Anthropic, \texttt{Qwen-QWQ-32B} from Alibaba~\citep{qwq32b}, and \texttt{deepseek-r1-distill-llama-70B} from DeepSeek~\citep{guo2025deepseek}. All models are used with a temperature of 0.0; other parameters remain at default settings.

To assess how well models identify and prioritize relevant information, we augment the \textit{retrieved context} setting by allowing each model to rerank the top-10 passages retrieved by \texttt{e5-mistral-7b-instruct}~\citep{wang2023improving} and select the top-5 most relevant ones. We then evaluate this step using the \texttt{Context Precision} metric with references, as provided by the RAGAS~\citep{es2023ragas} framework.

We consider three experimental settings to assess generative performance. In the \textit{no context} setting, models generate responses without any external information, simulating scenarios with no retrieval. In the \textit{retrieved context} setting, models are provided the top-10 retrieved passages. In the \textit{gold context} setting, models are given expert-annotated reference information, representing an ideal retrieval case. Prompts are kept minimal across all settings, presenting the context (if any) followed by the user query.
Generation quality is evaluated using \texttt{Correctness} and \texttt{Faithfulness}~\citep{es2023ragas}. \texttt{Correctness} measures factual and semantic alignment with the ground truth answer, while \texttt{Faithfulness} assesses consistency with the provided context. A response is considered faithful if all claims are supported by the context. Both metrics range from 0 to 100, with higher values indicating better performance.

\section{Results on \textsc{FinDER}}

We first evaluate retrieval models, highlighting how neural methods outperform traditional approaches in capturing domain-specific semantics. We then examine reranking with LLMs to improve contextual relevance. Finally, we assess generation models across diverse financial reasoning tasks and analyze how contextual grounding affects accuracy and faithfulness. Together, these results emphasize the importance of robust retrieval, effective reranking, and context-aware generation for financial QA in \textsc{FinDER}.

\subsection{Retrieval Performance Across Financial Domains}

\begin{table}[t]
\centering
\caption{The decoder-based retrieval model (E5-mistral) demonstrates the best performance in all categories in terms of Context Recall, while encoder-based models generally outperform BM25.}
\label{tab:retrieval_recall}
\begin{tabular}{lcccc}
\toprule
\textbf{Category} 
& \textbf{BM25} 
& \textbf{GTE} 
& \textbf{mE5} 
& \textbf{E5-mistral} \\
\midrule
Accounting           & 15.14 & 13.78 & 18.23 & \textbf{31.92} \\
Company overview     & 13.83 & 24.76 & 24.57 & \textbf{32.48} \\
Financials           & 6.42  & 11.92 & 9.14  & \textbf{15.84} \\
Footnotes            & 10.30 & 13.92 & 13.11 & \textbf{22.58} \\
Governance           & 8.57  & 14.16 & 13.49 & \textbf{19.11} \\
Legal                & 13.17 & 18.86 & 18.58 & \textbf{29.71} \\
Risk                 & 14.36 & 23.61 & 23.97 & \textbf{33.07} \\
Shareholder return   & 17.23 & 24.67 & 23.25 & \textbf{31.67} \\
Total                & 11.68 & 17.83 & 17.36 & \textbf{25.95} \\
\bottomrule
\end{tabular}
\end{table}

Table~\ref{tab:retrieval_recall} provides a comparison of retrieval performance across eight diverse financial document categories. While classical methods like BM25 rely on a traditional bag-of-words approach, neural models such as GTE, me5, and especially the decoder-based E5-mistral, provide richer, context-sensitive embeddings. The standout finding here is that E5-mistral significantly surpasses other methods, consistently demonstrating superior Context Recall. Notably, neural embedding models, regardless of architecture, uniformly outperform BM25. This underlines the transformative impact of learned semantic representations in capturing nuanced financial domain knowledge.

\begin{table}[t]
\centering
\caption{Comparison of retrieval performance between well-formed questions and \textsc{FinDER} using Precision. Well-formed questions are manually rewritten by financial experts to expand domain-specific terminology for a random sample of 500 queries within \textsc{FinDER}.}
\label{tab:retrieval_compare_wellformed}
\begin{tabular}{lcc}
\toprule
\textbf{Models} & \textbf{Well-formed Questions} & \textbf{\textsc{FinDER}} \\
\midrule
BM25 & \textbf{13.1} & 10.8 \\
GTE & \textbf{20.2} & 18.1 \\
mE5 & \textbf{21.0} & 17.5 \\
E5-Mistral & \textbf{33.9} & 25.7 \\
\bottomrule
\end{tabular}
\end{table}

Table~\ref{tab:retrieval_compare_wellformed} brings into sharp relief the critical role of query quality. The analysis compares performance between \textit{well-formed queries}, carefully refined by financial domain experts to remove ambiguity, and real-world queries from \textsc{FinDER}. The gap in precision highlights a fundamental insight: real-world queries often suffer from brevity and ambiguity, significantly challenging retrieval performance. Thus, \textsc{FinDER}'s real-world complexity offers researchers a valuable benchmark, spotlighting the pressing need for retrieval models that robustly interpret ambiguous, domain-specific queries.

\subsection{Reranking with Language Models}

\begin{table*}[t]
\centering
\caption{F1-score evaluation of reranking performance, where each model reranks the top 10 retrieved results from \texttt{e5-mistral-7b-instruct}~\citep{wang2023improving} and selects the top 5.
Recall is computed as the proportion of ground-truth answer elements correctly attributed to any of the retrieved contexts, while precision is calculated as the average precision based on the order of predicted relevant contexts~\citep{es2023ragas}.
}
\label{tab:reranking}
\resizebox{0.9\textwidth}{!}{%
\begin{tabular}{lcccc}
\toprule
\textbf{Category} 
& \textbf{Claude-3.7-Sonnet} 
& \textbf{GPT-o1} 
& \textbf{Deepseek-R1-Distill} 
& \textbf{Qwen-QWQ} \\
\midrule
Accounting         & 79.37 & 82.62 & \textbf{84.71} & \underline{83.33} \\
Company overview   & \textbf{63.29} & \underline{62.27} & 59.58 & 59.99 \\
Financials         & \underline{50.04} & \textbf{50.46} & 43.75 & 47.36 \\
Footnotes          & \textbf{63.03} & \underline{61.59} & 59.29 & 60.53 \\
Governance         & \underline{53.37} & \textbf{53.77} & 50.40 & 51.16 \\
Legal              & \underline{77.11} & 76.00 & 71.47 & \textbf{80.22} \\
Risk               & \underline{79.90} & \underline{79.90} & 78.27 & \textbf{80.32} \\
Shareholder return & \textbf{45.91} & \underline{43.15} & 41.41 & 42.36 \\
\midrule
\textbf{Total}     & \textbf{63.05} & \underline{62.90} & 60.01 & 61.78 \\
\bottomrule
\end{tabular}
}
\end{table*}

In Table~\ref{tab:reranking}, we delve deeper into the subtle yet crucial art of context reranking, evaluated by F1-score. Models rerank the top 10 contexts initially retrieved by e5-mistral-7b-instruct, selecting the five most relevant. Here, large language models (LLMs) like Claude-3.7-Sonnet and GPT-o1 clearly shine, consistently achieving high performance across various financial categories, indicating their superior reasoning and context discernment capabilities. Interestingly, more specialized models such as Deepseek-R1-Distill exhibit notable category-specific strengths, particularly in Accounting, whereas Qwen-QWQ excels in Legal and Risk domains. These nuanced performances suggest a critical insight: retrieval sets need not be perfectly precise—rather, a diverse retrieval pool is beneficial since reasoning-focused models effectively discern relevant information despite some noise.

\subsection{Generation Under Financial Reasoning Tasks}

\begin{table*}[t]
\centering
\caption{Comparison of four baseline language models across six tasks (Qualitative, Addition, Subtraction, Multiplication, Division, Compositional, and Total), evaluated by Response Correctness and Faithfulness metrics. The experiment is conducted in the setting of \textit{Using partial information}, where only the top-10 retrieval results from \texttt{e5-mistral-7b-instruct}~\citep{wang2023improving} are provided as context.}
\label{tab:generation_by_task}
\resizebox{0.9\textwidth}{!}{%
\begin{tabular}{lcccccccc}
\toprule
\multirow{2}{*}{\textbf{Task}} 
& \multicolumn{2}{c}{\textbf{Claude-3.7-Sonnet}} 
& \multicolumn{2}{c}{\textbf{GPT-o1}} 
& \multicolumn{2}{c}{\textbf{Deepseek-R1-distill}} 
& \multicolumn{2}{c}{\textbf{Qwen-QWQ}} \\
\cmidrule(lr){2-3} 
\cmidrule(lr){4-5} 
\cmidrule(lr){6-7}
\cmidrule(lr){8-9}
& \textbf{Corr.} & \textbf{Faith.}
& \textbf{Corr.} & \textbf{Faith.}
& \textbf{Corr.} & \textbf{Faith.}
& \textbf{Corr.} & \textbf{Faith.} \\
\midrule
\textbf{Qualitative}        & 30.06 & \textbf{85.46} & \underline{33.28} & \underline{84.15} & 32.57 & 75.74 & \textbf{34.11} & 81.93 \\
\textbf{Quantitative}          & 22.82 & \textbf{81.66} & \textbf{25.24} & \underline{79.05} & 23.32 & 70.96 & \underline{24.06} & 70.46 \\
\quad $\llcorner$ Addition              & 18.61 & \underline{77.38} & \underline{20.21} & 71.98 & \textbf{21.67} & 73.20 & 15.64 & \textbf{77.77} \\
\quad $\llcorner$ Subtract              & 19.88 & \textbf{86.11} & \textbf{28.76} & \underline{80.65} & 24.31 & 76.55 & \underline{24.55} & 74.01 \\
\quad $\llcorner$ Multiplication        & 34.87 & \underline{79.36} & \textbf{42.90} & \textbf{81.44} & 33.00 & 49.89 & \underline{36.33} & 61.26 \\
\quad $\llcorner$ Division              & 27.49 & \textbf{83.98} & 27.78 & \underline{80.33} & \underline{28.24} & 71.43 & \textbf{31.69} & 69.82 \\
\quad $\llcorner$ Composition           & \textbf{20.40} & \textbf{80.67} & \underline{19.93} & \underline{79.13} & 19.36 & 72.54 & 19.64 & 69.62 \\
\textbf{Total}              & 28.79 & \textbf{84.75} & \underline{31.89} & \underline{83.35} & 30.96 & 74.89 & \textbf{32.41} & 79.99 \\
\bottomrule
\end{tabular}
}
\end{table*}

Exploring further, Table~\ref{tab:generation_by_task} assesses baseline language models across seven tasks—Qualitative reasoning and various arithmetic operations (Addition, Subtraction, Multiplication, Division, Composition)—evaluated through metrics such as Response Correctness and Faithfulness. The results reveal intriguing nuances: GPT-o1 and Qwen-QWQ excel in arithmetic tasks such as Multiplication and Division, showcasing impressive numeric reasoning capabilities. Meanwhile, Claude-3.7-Sonnet consistently achieves the highest Faithfulness scores, highlighting its unique ability to produce coherent and trustworthy outputs. However, a clear insight emerges—no single model universally dominates, underscoring that architectural nuances significantly shape model strengths across specific reasoning tasks.

Finally, Table~\ref{tab:generation_by_info} examines how varying levels of contextual information affect model performance. This analysis starkly illustrates that providing richer context dramatically enhances response correctness. Specifically, under the Perfect Information scenario—where models receive precise, relevant financial context—models like Claude-3.7-Sonnet and GPT-o1 display the most pronounced improvements. The central lesson here is powerful yet straightforward: accurate and relevant context provision is vital for robust and meaningful generation in financial applications, emphasizing the intertwined importance of retrieval effectiveness and contextual grounding for model success.

\begin{table*}[t]
\centering
\caption{Evaluation of four baseline LLMs under three context conditions—\textit{without context}, \textit{partial context}, and \textit{perfect context}—based on response correctness. 
In the \textit{without context} setting, no external context is provided. 
The \textit{partial context} setting uses the top-10 retrieved results from \texttt{e5-mistral-7b-instruct}~\citep{wang2023improving} as context. 
The \textit{perfect context} setting provides a section of the document that contains the ground-truth context relevant to answering the question.}
\label{tab:generation_by_info}
\resizebox{0.9\textwidth}{!}{%
\begin{tabular}{lcccc}
\toprule
\textbf{Information Setting} 
& \textbf{Claude-3.7-Sonnet} 
& \textbf{GPT-o1} 
& \textbf{Deepseek-R1-Distill} 
& \textbf{Qwen-QWQ} \\
\midrule
\textbf{Without Information} & 9.37 & \textbf{10.14} & \underline{9.88} & 9.13 \\
\textbf{Using Top-10 Retrieved Results} & \textbf{33.89} & \underline{32.96} & 28.79 & 29.41 \\
\textbf{Perfect Information} & \underline{66.48} & \textbf{68.13} & 59.69 & 61.03 \\
\bottomrule
\end{tabular}
}
\end{table*}

Our findings highlight that robust financial QA requires more than strong generation models. Effective retrieval and reranking play a central role, especially under noisy real-world inputs. Neural retrievers like E5-mistral set a strong foundation, and LLM-based reranking compensates for imperfect retrieval. Ultimately, grounding generation in high-quality context is key to delivering accurate, faithful answers across diverse financial tasks.

\section{Conclusion}

\textsc{FinDER} establishes a new benchmark for financial question-answering by introducing ambiguous, domain-specific queries that reflect real-world search behavior. It challenges models to retrieve and synthesize relevant information from expert-annotated financial documents, offering a more realistic evaluation framework for Retrieval-Augmented Generation (RAG). Our results show that dense retrieval model like e5-mistral outperforms traditional sparse methods but still struggles with ambiguous queries, highlighting the need for improved retrieval strategies. In generation tasks, models perform significantly better with high-quality retrieved context, yet no single model consistently excels across all financial reasoning tasks. \textsc{FinDER} underscores the gap between current AI capabilities and the precision required in finance, providing a rigorous testbed for advancing retrieval algorithms, query disambiguation, and factually accurate text generation. Future work should explore integrating diverse financial document sources and developing retrieval-enhanced models that refine queries dynamically.

\bibliography{iclr2025_conference}
\bibliographystyle{iclr2025_conference}

\end{document}